\documentclass[aps,prb,showpacs,twocolumn,10pt,longbibliography]{revtex4-1}

\usepackage{graphicx}
\usepackage{amssymb}
\usepackage{amsmath}
\usepackage{color}
\usepackage{bm}
\usepackage{verbatim}

\usepackage{amsfonts}
\usepackage{graphicx}
\usepackage{bm}
\usepackage{color}
\usepackage{epstopdf}

\begin{document}

%%%%%%%%%%%%%%%%%%%%%TITLE%%%%%%%%%%%%%%%%%%%%%%%%%%%%%%%%%%%%%%%%%%%%%%%%%%%%%%%%%%%
\title{High temperature magnetism and microstructure of semiconducting ferromagnetic alloy (GaSb)$_{1-x}$(MnSb)$_x$}

\author{
L.N. Oveshnikov$^{1,2,*}$, E.I. Nekhaeva$^{1,2}$, A.V. Kochura$^{3}$, A.B. Davydov$^2$, M.A. Shakhov$^{4,5}$, S.F. Marenkin$^6$, O.A. Novodvorskii$^7$, A.P. Kuzmenko$^{3}$, A.L.Vasiliev$^1$, B.A. Aronzon$^{1,2}$ and E. Lahderanta$^4$
}

\affiliation{$^1$ National Research Center "Kurchatov Institute", 123182 Moscow, Russian Federation}
\affiliation{$^2$ P.N. Lebedev Physical Institute, Russian Academy of Sciences, 119991 Moscow, Russian Federation}
\affiliation{$^3$ South-West State University, 305040 Kursk, Russian Federation}
\affiliation{$^4$ Lappeenranta University of Technology, 53850 Lappeenranta, Finland}
\affiliation{$^5$ Ioffe Institute, 194021 St.Petersburg, Russian Federation}
\affiliation{$^6$ Kurnakov Institute of General and Inorganic Chemistry, Russian Academy of Sciences, 119991 Moscow, Russian Federation}
\affiliation{$^7$ Institute on Laser and Information Technologies, Russian Academy of Sciences, 140700 Shatura, Moscow Region, Russian Federation}

%%%%%%%%%%%%%%%%%%%%%%%%%%ABSTRACT%%%%%%%%%%%%%%%%%%%%%%%%%%%%%%%%%%%%%%%%%%%%%%%%%%%%%%%%%%%%%%%%%%%%%%%%%%%%
\begin{abstract}
 We have studied the properties of relatively thick (about 120 nm) magnetic composite films grown by pulsed laser deposition method using (GaSb)$_{0.59}$(MnSb)$_{0.41}$ eutectic compound as a target for sputtering. For the studied films we have observed ferromagnetism and
 anomalous Hall effect above the room temperature, it manifests the presence of spin-polarized carriers. Electron microscopy, atomic and magnetic force microscopy results suggests that films under study have homogenous columnar structure in the bulk while MnSb inclusions accumulate near it's surface. This is in good agreement with high mobility values of charge carriers. Based on our data we conclude that room temperature magnetic and magnetotransport properties of the films are defined by MnSb inclusions.
\end{abstract}
%%%%%%%%%%%%%%%%%%%%%%%%%%%%%%%%%%%%%%%%%%%%%%%%%%%%%%%%%%%%%%%%%%%%%%%%%%%%%%%%%%%%%%%%%%%%%%%%%%%%%%%%%%%%%%%%

%\pacs{72.10.-d, 72.15.Lh, 72.80.Vp, 72.80.Ng}

%%%%%%%%%%%%%%%%%%%%%%%%%%%%%%%%%%%%%%%%%%%%%%%%%%%%%%%%%%
%71.55.Ak	Metals, semimetals, and alloys
%72.10.-d	Theory of electronic transport; scattering mechanisms
%78.40.Kc	Metals, semimetals, and alloys
%72.15.Lh	Relaxation times and mean free paths (in metals and alloys)
%72.80.Ng	Disordered solids (under transport in specific materials)
%72.80.Vp	Electronic transport in graphene

%\date{\today}

\maketitle

%%%%%%%%%%%%%%%%%%%%%%%%%%%%%%%%%%%%%%%%%%%%%%%%%%%%%%%%%%%%%%%%%%%%%%%%%%%%%%%%%%%%%
\section{Introduction}
The diluted magnetic semiconductors (DMS) are very promising materials for spintronic devices, because DMS offer the combination of magnetic and semiconducting properties. Currently the most widely studied DMS systems are those based on III-V semiconductors doped by Mn~\cite{jung2006,dietl2007,dietl2014}. Among these systems  the most known and better studied is Ga$_{1-x}$Mn$_x$As. Here Mn substitute Ga atoms and establish a ferromagnetic state realized via carrier-induced indirect exchange between Mn atoms by Zener - RKKY mechanism accompanied by the spin polarization of conducting holes~\cite{jung2006,dietl2014}. To reach high $T_c$ values one needs material with high Mn concentration, which can be achieved by using non-equilibrium growth methods, such as low-temperature molecular-beam epitaxy. In particular, it increases the solubility limit of Mn in Ga$_{1-x}$Mn$_x$As up to $x = 0.2$ without precipitate formation ~\cite{chen2009}. The highest Curie temperature $T_c$ achieved in these materials was below 200 K, observed with $x$ about 0.1 ~\cite{wang2014,chen2011}. This is remarkably high for a DMS system while for practical applications is desired $T_c>300$ K. At higher concentrations Mn atoms start to occupy interstitial sites and produce strong structure defects that increase scattering of charge carriers. Thus, hole mobilities in Ga$_{1-x}$Mn$_x$As systems with high $T_c$ ($x> 0.06$) usually do not exceed 10 cm$^2/$(V$\cdot$s).

Another way to realize high $T_c$ in magnetic semiconductor material is to create a granular system with two phases, i.e. ferromagnetic nanoinclusions embedded into semiconductor matrix. Although, this type of systems are studied less often, in some of the related works observed $T_c$ exceeded room temperature ~\cite{lawnPSS2011,lawnJAP2011,rylkov2005,braun2007,maren2014,kopl2015,talan2016}. Additional advantage of granular systems is higher values of carrier mobility, about one order of magnitude higher than that in traditional DMS such as Ga$_{1-x}$Mn$_x$As. That is due to the aggregation of the most part of magnetic impurity atoms within nanoinclusions, which results in higher crystalline quality of semiconductor matrix~\cite{rylkov2005} and lower density of scatterers. Thus, granular materials could be of interest, both as an object of fundamental studies of DMS systems and as a versatile material suitable for testing prototype spintronic devices.

Recent studies of MnAs inclusions embedded into GaAs matrix showed that inclusions can emerge with two types of crystal structure. Furthermore, the magnetic properties of MnAs inclusions with zinc blende type and hexagonal lattices are substantially different. That is why the actual application of GaAs:MnAs material is restricted by this obstacle.

Second recently studied nanocomposite system is GaSb matrix with incorporated MnSb nanograins~\cite{lawnJAP2011,rylkov2005,braun2007,mats2000}. However, the best results were obtained not for a composite system but for the (GaSb)$_{1-x}$(MnSb)$_x$ alloys with $x=0.41$. In particular, for annealed samples the mobility of holes was about 100 cm$^2/$(V$\cdot$s) and the $T_c$ was above room temperature~\cite{maren2014,kopl2015}. Earlier it was suggested~\cite{rylkov2005} that the ferromagnetic ordering in this case is induced by interaction of MnSb magnetic clusters with carriers inside the matrix. It should induce carrier spin-polarization and lead to the formation of long range ferromagnetic percolation cluster which includes both MnSb magnetic clusters and spin-polarized carriers. However, to check this assumption and to reveal the origin of the high temperature ferromagnetism in GaSb-MnSb alloys, one needs a detailed knowledge of the sample structure.

Thus, in this paper we investigate magnetic and transport properties of the GaSb-MnSb alloy films with $x=0.41$ and we elucidate the origin of the ferromagnetic state in this material. We use Atomic Force and Magnetic Force Microscopy (AFM and MFM) as well as scanning/transmission electron microscopy (S/TEM) to study the samples structure.

\section{Samples and Methods}
GaSb-MnSb films with thickness ($d$) in the range between 120 nm and 135 nm and an area of $0.1-1$ cm$^2$ were grown by the droplet-free pulsed laser deposition (PLD) method in a high vacuum of 10$^{-6}$ Torr with deposition temperatures $T_{dep}=100-300$ $^\circ$C. We used GaSb-MnSb eutectic composition containing 41 mol.$\%$ MnSb and 59 mol.$\%$ GaSb as a target, that was sputtered by the second harmonic radiation of a Q-switched YAG:Nd laser ($\lambda = 532$ nm). Al$_2$O$_3$ (0001) single crystals were used as substrates. More detailed description of the growth technology can be found in~\cite{lotin2010}.

Magnetization was measured at temperatures $T=5-310$ K in magnetic fields up to $H=50$ kOe using a superconducting quantum interference device (SQUID) magnetometer S600X (Cryogenic, UK). The electrical and magnetotransport properties were investigated at temperatures $T=2-320$ K using standard six-probe geometry in pulsed magnetic fields up to $H=300$ kOe. Studied samples demonstrated linear current-voltage characteristics down to sub-helium temperatures while sustaining high values of conductivity.

The cross-section specimens for S/TEM studies was prepared by focus ion beam (FIB) milling procedure in a Helios (FEI, US) scanning electron microscope (SEM)/FIB dual beam system equipped with C and Pt contain gas injectors and a micromanipulator (Omniprobe, US). A 2 $\mu$m Pt layer was deposited on the surface of the sample prior to the cross-sections preparation by FIB milling procedure. Sections of approximately 8$\times$5 $\mu$m$^2$ in size and 2 $\mu$m thick were cut by 30 kV Ga+ ions, removed from the sample and then attached to the Omniprobe semiring (Omniprobe, US). Final thinning was performed with 5 kV Ga$^+$ ions followed by cleaning by 2 keV Ga$^+$ ions for the electron transparency. All specimens were studied in a scanning/transmission electron microscope (S/TEM) Titan 80-300 (FEI, US) equipped with a spherical aberration (Cs) corrector (electron probe corrector), a high-angle annular dark field (HAADF) detector, an atmospheric thin-window energy dispersive x-ray (EDX) spectrometer (Phoenix System, EDAX, US) and a post-column Gatan energy filter (GIF), (Gatan, US). The S/TEM was operated at 300 kV. Digital micrograph (Gatan, US) and TIA software (FEI, US) was used for image analysis. P.Stadelmann's JEMS software~\cite{stad1987} was used for diffraction patterns and image simulation.

Scanning atomic force microscopy (AFM) and magnetic force microscopy (MFM) images were obtained on an SmartSPM microscope (AIST-NT, US)  at temperatures $T=295-450$ K.

\section{Results and Discussion}

\begin{table}[t]
\caption{Sample parameters: deposition temperature $T_{dep}$; film thickness $d$; carrier concentration $p$; carrier mobility $\mu$; coercive force $H_c$; remanent magnetization $M_{rem}$; saturation magnetization $M_{sat}$ (\emph{$p$, $\mu$ and $M_{sat}$ were obtained at $T=300$K, while $H_c$ and $M_{rem}$ values correspond to $T=4.2$K}).}
\begin{center}
\begin{tabular}{|c|*{10}{c|}}
%%{ccc}
%%Sample $\emph{\textbf{x}}$ & $\emph{\textbf{d}}$, & \mbox{\boldmath$\mu$}, & $\emph{\textbf{n}}$, & $\emph{\textbf{n}}$, & $\emph{\textbf{n}}$, & $\emph{\textbf{n}}$, & $\emph{\textbf{n}}$,  \\
\hline
\textbf{Sample} & $T_{dep}$, &  $d$, & $p$, & $\mu$, & $H_c$,$^*$ & $M_{rem}$$^*$ & $M_{sat}$$^*$ \\
  & $^\circ$C & nm & 10$^{19}$ cm$^{-3}$ & cm$^2$/$(V\cdot s$) & Oe & &   \\
\hline
 GM1 & 100 & 120 & 0.8 & 110 & 250 & 1.4 $\mu_B$ & 1.8 $\mu_B$ \\
\hline
 GM2 & 100 & 120 & 9.3 & 110 & 260 & 2.1 $\mu_B$ & 3.0 $\mu_B$ \\
\hline
 GM3 & 200 & 130 & 23 & 80 & 260 & 1.7 $\mu_B$ & 3.0 $\mu_B$ \\
\hline
 GM4 & 200 & 135 & 14 & 71 & 115 & 2.1 $\mu_B$ & 3.2 $\mu_B$ \\
\hline
 GM5 & 300 & 120 & 4.8 & 74 & 265 & 0.6 $\mu_B$ & 2.5 $\mu_B$ \\
\hline
\end{tabular}
\end{center}
 $^*$\emph{Magnetic parameters were obtained for magnetic field oriented parallel to the sample plain; magnetization were calculated per manganese atom}.
\end{table}

We have studied several (GaSb)$_{1-x}$(MnSb)$_x$ samples with $x=0.41$, both annealed and not annealed. In this paper we discuss only samples annealed at 350 $^\circ$C during 30 min with high holes concentration $p$ ($> 10^{19}$ cm$^{-3}$) because they showed much better magnetic and semiconducting (electron transport) properties. Therefore, these samples are more suitable to reveal the nature of magnetic properties and hole spin-polarization in this material which is the aim of this paper. Main parameters of the studied samples are presented in the Table 1.

Fig. 1a shows typical curves of magnetization versus magnetic field for the sample GM3 (see Table 1) for magnetic field orientations in the sample plane and perpendicularly to it. The presence of well-pronounced hysteresis suggests that ferromagnetic ordering in these materials appears at $T>300$ K. This is more evident from the temperature dependence of the remanent magnetization $M_{rem}$ presented in Fig. 1b. At the same figure we also provide the MFM data and temperature dependence of the remanent magnetization for InSb-MnSb eutectics. Nonzero $M_{rem}$ persists up to 400 K and it's temperature dependence is close to that for InSb-MnSb eutectic composition which corresponds to the MnSb $T_c\approx$ 600 K~\cite{novo2011}. Thus, presented data clearly proves the presence of high temperature ferromagnetism in the samples under study.

\begin{figure}
\centering
\includegraphics[width=8cm,clip]{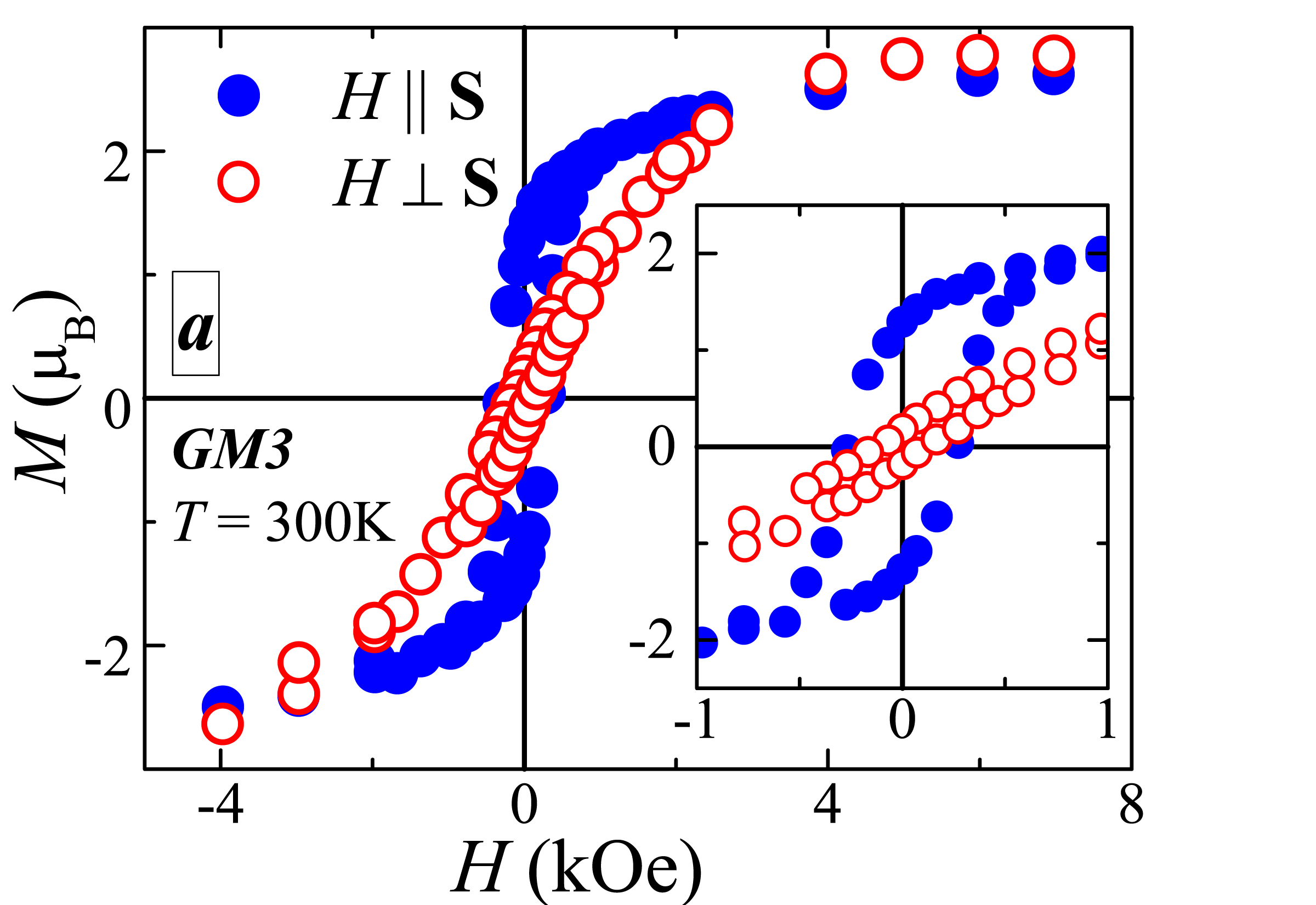}
\includegraphics[width=8cm,clip]{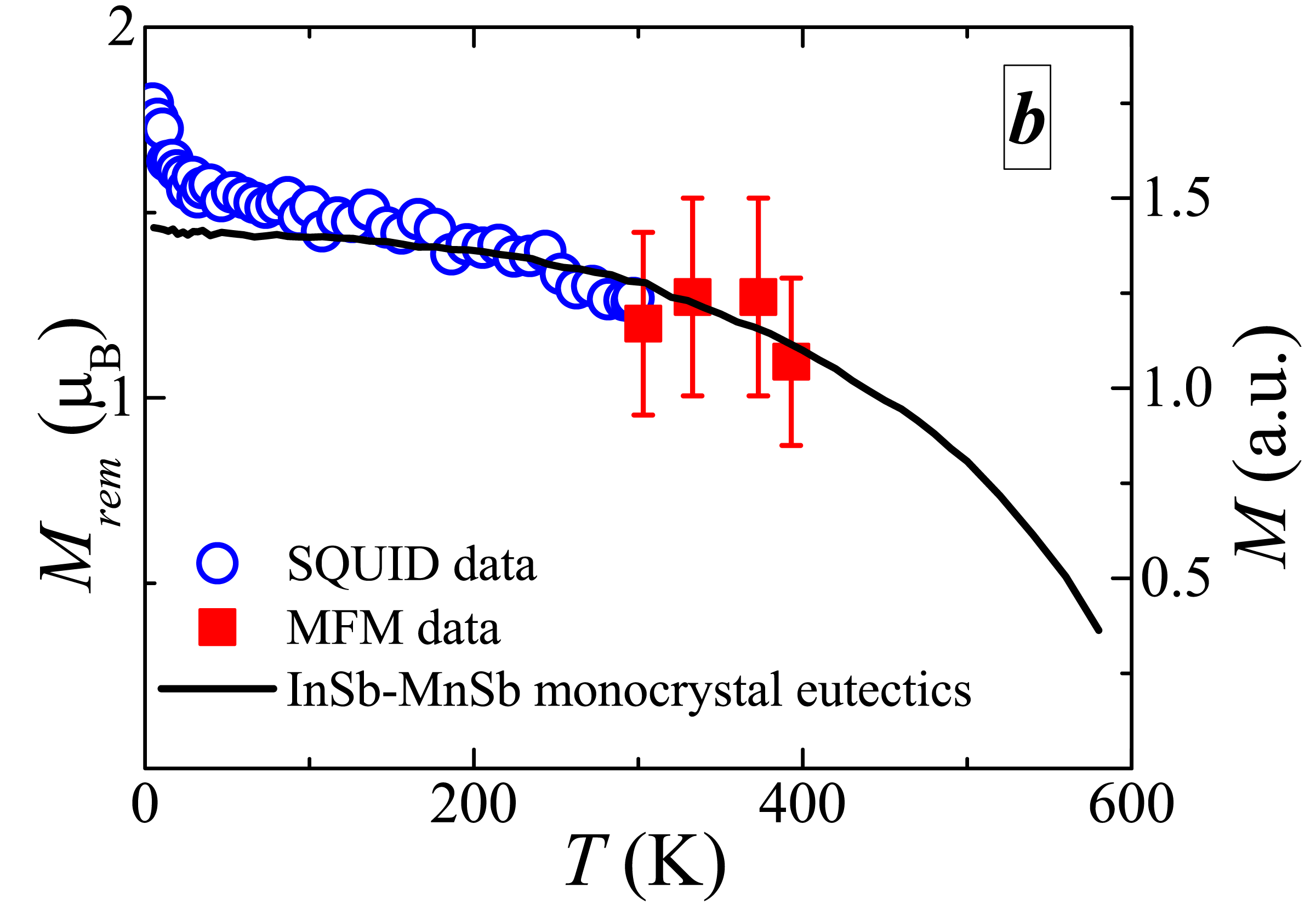}
\caption{(colour online) (a) Magnetization vs. magnetic field for sample GM3 at $T=300$ K. Measurements were performed with magnetic field oriented in the sample plane (solid symbols) and perpendicular to it (open symbols). Inset demonstrate hysteresis loop at low fields. (b) Temperature dependence of the remanent magnetization. Open circles are SQUID data for sample GM3, red squares are MFM data for the same sample and black curve is temperature dependence of the saturation magnetization of InSb-MnSb eutectics~\cite{novo2011}.}
\label{fig1}
\end{figure}

As said earlier, studied systems differs from traditional DMS materials by substantially higher carrier mobilities and $T_c$ values, while the nature of high temperature ferromagnetism is not completely clear. Previous studies~\cite{rylkov2005} suggested that ferromagnetism in (GaSb)$_{1-x}$(MnSb)$_x$ is related to the interaction of charge carriers with MnSb nanoinclusions with $T_c=600$ K~\cite{novo2011,tera1968}. This interaction is affected by the appearance of Shottky barriers at the MnSb/GaSb boundaries. Basically, these type of barriers appear on the semiconductor/metal interfaces providing a tunneling charge transfer across the boundary, if the barrier is high enough. In the present case Shottky barriers may appear only if inclusions are sufficiently large to establish second electronic phase (metal-type) despite the impact of surrounding GaSb matrix. Parameters of these barriers (e.g. the width) substantially depend on charge distribution, i.e. they depend on carrier concentration.

It is worth mentioning that in real samples not all Mn atoms are strictly positioned in MnSb inclusions, some part of Mn atoms can be distributed within semiconductor matrix. It results in two-phase magnetic subsystem for which lower $T_c$ corresponds to isolated Mn atoms within matrix~\cite{M6}. Also, as it is shown in Table 1, the $M_{sat}$ values reach 3.2 $\mu_B$ per Mn atom which is close to 3.6 $\mu_B$ value obtained from the experimental data for MnSb samples analyzed earlier~\cite{tera1968,coeh1985}, but is lower than expected for Mn$^{2+}$. This difference can be related to the presence of Mn$^{3+}$ ions and antiferromagnetic character of interaction between carriers and Mn atoms.

\begin{figure}
\centering
\includegraphics[width=8.5cm]{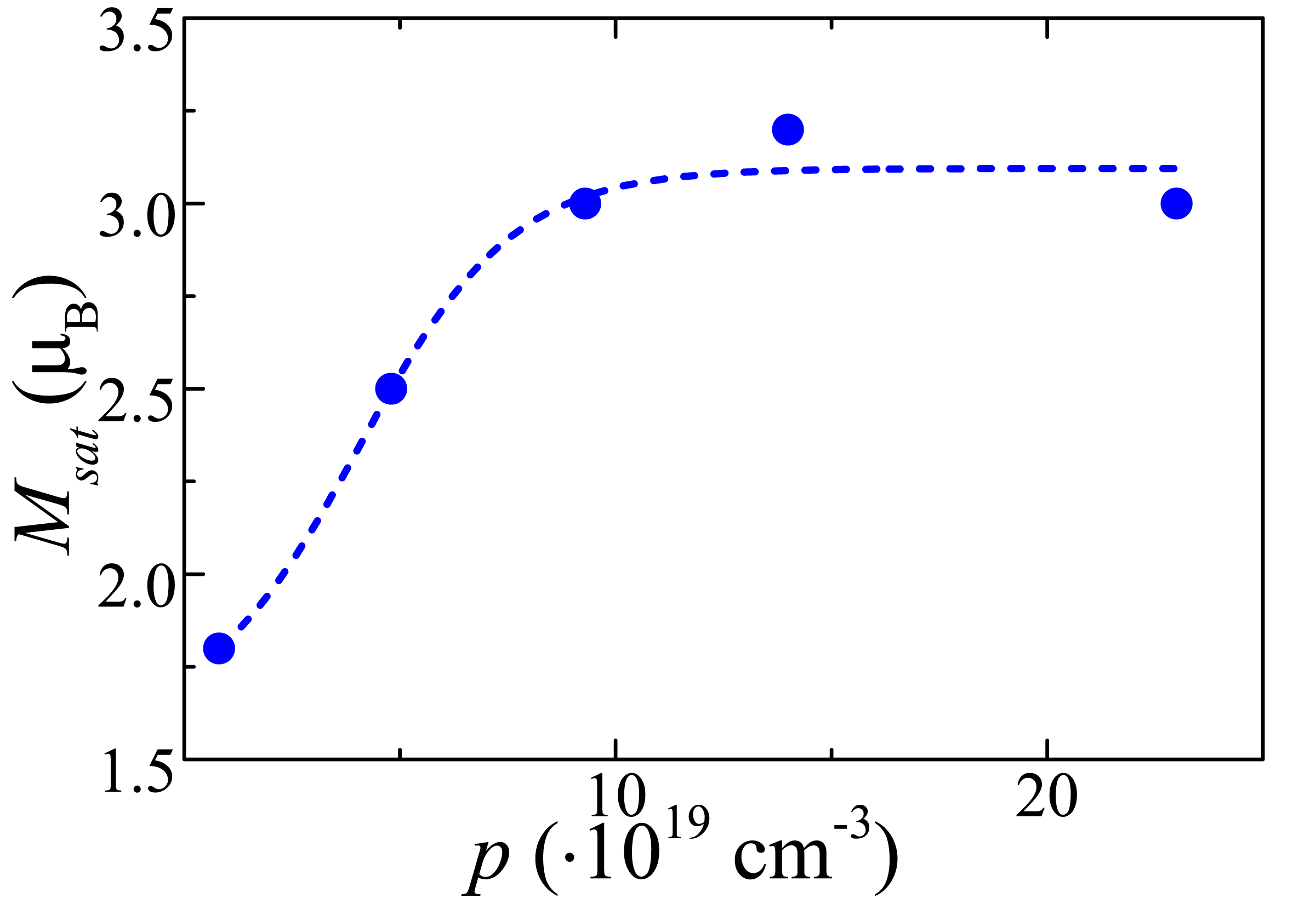}
\caption{(colour online) Room temperature saturation magnetization ($M_{sat}$) versus carrier concentration ($p$) dependence (dash line is a guide to the eye).}
\label{fig2}
\end{figure}

One major parameter of holes interaction with MnSb inclusions is the width of Shottky barriers $d_{barrier}$ which decreases with the growth of carrier concentration~~\cite{rylkov2005}. This picture correlates with concentration dependence of saturation magnetization $M_{sat}$ presented in Fig. 2. This also agrees well with the data obtained for previously studied (GaSb)$_{1-x}$(MnSb)$_x$ films (see Fig. 5 in~\cite{kopl2015}). As it can be seen from these figures, $M_{sat}$ increases with carrier concentration and saturates above $p\approx 10^{20}$ cm$^{-3}$ at which $d_{barrier}$ becomes comparable with the effective penetration depth of carrier (hole) wave function under triangular barrier $l_p$. Taking into account that energy gap in GaSb is $E_g=0.7$ eV and the Schottky barrier height is about $1/3$ $E_g$~\cite{sze1981}, the $d_{barrier}$ can be estimated as 2 nm at $p=10^{20}$ cm$^{-3}$ while $l_p$ is of the same value~\cite{rylkov2005}.
As shown in Fig. 2, $p$ values for studied samples are sufficiently high to provide effective holes interaction with MnSb inclusions and cause the saturation of magnetization. This, along with high $T_c$ values, suggests that this interaction could be the main source of high temperature ferromagnetism in studied systems. However, to justify the applicability of carrier-to-inclusions interaction model one need to verify specific properties of spin-polarized system, e.g. displayed in transport measurements.

\begin{figure*}
\centering
\includegraphics[width=14cm]{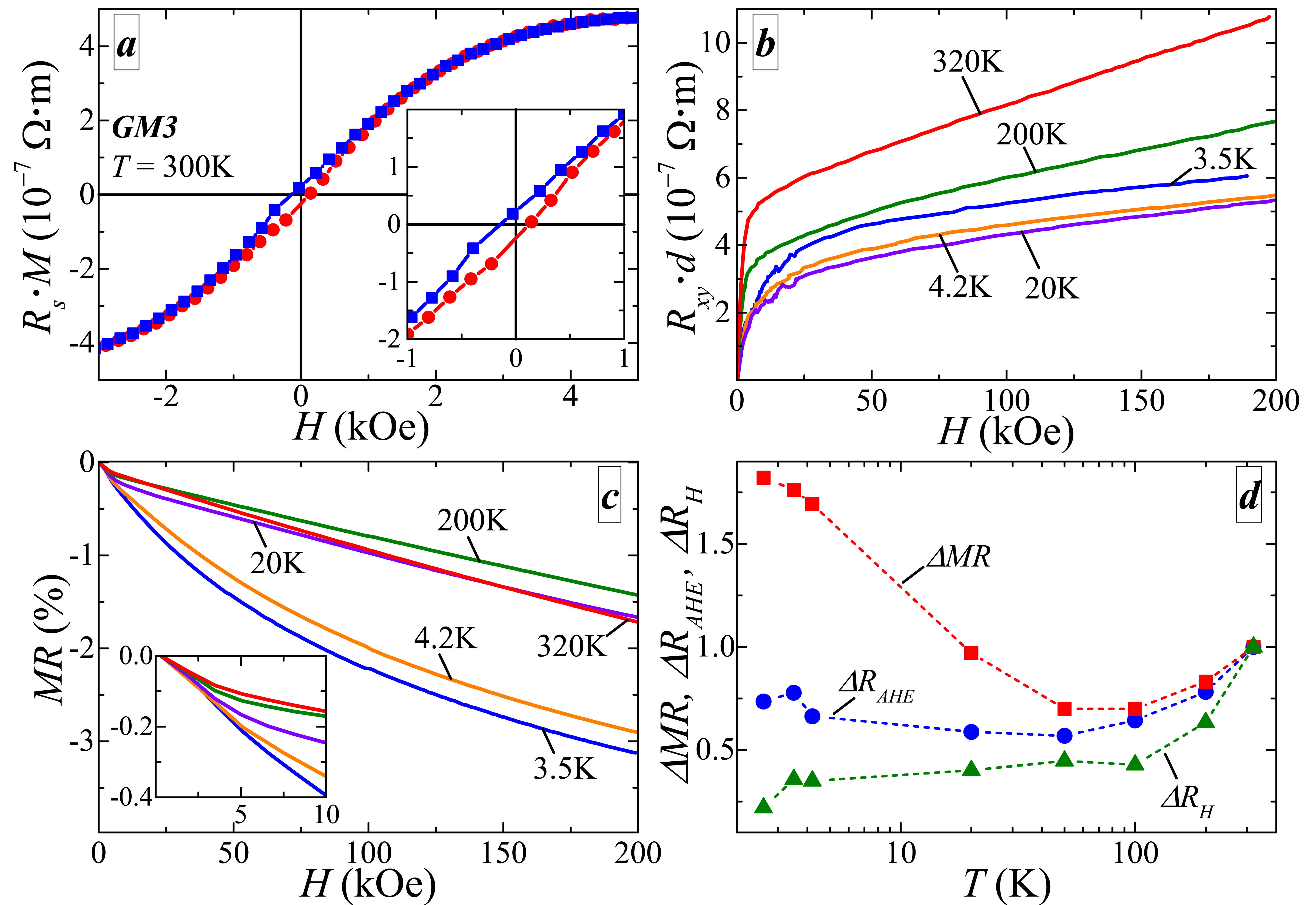}
\caption{(colour online) Magnetotransport properties of sample GM3: (a) Field dependence of anomalous Hall component at 300 K (here we have subtracted linear background). Slope of the curve correlates with the sign of charge carriers (holes), i.e. observed AHE is positive. Inset demonstrate hysteresis behaviour at low fields. (b) Field dependence of Hall resistance up to $H=200$ kOe at various temperatures. (c) Magnetoresistance at various temperatures. Inset shows low field part of presented curves. (d) Temperature dependence of high field magnetoresistance $\Delta MR$ (at $H=200$ kOe), saturated AHE amplitude $\Delta R_{AHE}$ and Hall slope $\Delta R_H$ normalized by corresponding values at $T=320$ K.
}
\label{fig3}
\end{figure*}

It should be noted, that mobility values have no pronounced dependence on carrier concentration (see Table 1), i.e. the conductivity of studied films is affected by various factors. Thus, magnetotransport phenomena can yield additional information on system properties. The results of Hall resistivity measurements for the GM3 sample are shown in Fig. 3a and 3b. The Hall resistivity in magnetic systems can be divided into two parts:
\begin{equation}
R_{xy}\cdot d=R_H{\bf H}+R_s{\bf M},\label{eq:1}
\end{equation}
where $R_H$ is Hall constant, used for $p$ and $\mu$ determination (see Table 1), and $R_s$ is anomalous Hall constant, which is defined by several parameters of the system and corresponding mechanism~\cite{nagaosa}. Thus, the  $R_{xy}(H)$ should reproduce field dependence of magnetization. As it shown in Fig. 3a $R_{s}\cdot M(H)$ (Hall resistivity after subtraction of linear contribution) demonstrates hysteresis behavior (see inset) with $H_c\approx140$ Oe and saturation field $H_{sat}\approx3.8$ kOe. This is in a good agreement with values $H_c\approx130$ Oe and $H_{sat}\approx4$ kOe obtained from SQUID data for $H$ perpendicular to the sample plain. Note, that $H_c$ and $H_{sat}$ values in Table 1 were obtained for $H$ oriented parallel to the sample plain, thus, the difference between them and parameters obtained from $R_{xy}(H)$ curves is related to substantial magnetic anisotropy of samples under study (see Fig. 1). The observation of anomalous Hall effect (AHE) clearly suggests that delocalized holes interact with magnetic subsystem, i.e. there are spin-polarized carriers even at room temperature. Thus, the exchange interaction within studied system can be mediated by these spin-polarized holes as it was suggested earlier.

The incorporation of experimentally observed high-temperature ferromagnetism and spin polarization of conducting carriers already makes studied films very perspective for various applications. But to get a deeper insight into properties of these films we have made additional magnetotransport measurements in high magnetic fields. Hall resistivity dependencies in magnetic fields up to $H=200$ kOe at various temperatures are presented in Fig. 3b. It is clearly seen that at room temperature the anomalous Hall effect contribution saturates below $H=10$ kOe and $R_{xy}(H)$ become linear. The linear slope and the saturation field are different for different temperatures. To visualise the temperature evolution of presented curves we used simple linear fit of high field region. In this case fitting function have two variables, slope ($R_H$) and an offset, which corresponds to saturated AHE amplitude $R_{AHE}$. Second common sign of interaction between magnetic and conducting subsystems is the appearance of negative magnetoresistance (nMR), which is usually ascribed to spin-dependent scattering. As it is shown in Fig. 3c studied samples demonstrate nMR which does not saturate up to 200 kOe in studied temperature interval. It should be noted, that at low field nMR have similar form at all temperatures (see inset in Fig. 3c), while at higher fields shape of MR curves changes. In particular, above 20 K nMR is linear above 20 kOe, while below 4.2 K the shape of nMR is close to sublinear or logarithmic, which is more common for spin-dependent scattering contribution~\cite{xiao1992,M12,M16}.

Based on the assumption of two-phase magnetic subsystem (MnSb inclusions and isolated Mn atoms) we can qualitatively describe temperature evolution of magnetotransport parameters presented in Fig. 3d. The character of $\Delta R_{AHE}$ and $\Delta MR$ dependencies is rather similar, thus, they should have the same nature. At 50 K we observe local minimum in both curves. The increase above 50 K can be related to the presence of Shottky barriers, because, if both AHE and nMR corresponds to the interaction with MnSb inclusions, then their amplitudes are defined by tunneling intensity which increases with temperature. But below 50 K another contribution becomes significant, the interaction with isolated Mn moments. From our data we cannot define corresponding $T_c$ accurately because this interaction can be significant even above the ordering temperature~\cite{M12,M16}, while the total magnetic moment of isolated Mn atoms can be substantially smaller than that of MnSb inclusions. Temperature dependence of $\Delta R_H$ is more complicated. But in the present case of large AHE contribution, MR and AHE itself can can strongly affect $R_H$ through the relations of conductivity and resistivity tensors~\cite{shen2008}. It should be noted that nMR at high temperatures can be also due to tensor relations, although mentioned difference of nMR functional form at various temperatures suggests that, at least at low temperatures, nMR should be related to the spin-dependent scattering. Also there are several other phenomena that can be relevant. Thus, to elucidate their contributions a more thoughtful study with quantitative simulations is needed, but it is out of scope of present paper.

To establish high temperature ferromagnetism of MnSb inclusions via spin-polarized carriers their interaction have to be sufficiently strong. It implies that there have to be a large amount of such inclusions with distances between them less than carrier spin relaxation length. However, the comparatively high carrier mobilities in studied samples still leave some doubt on the mentioned idea, because big amount of MnSb inclusions (as well as high concentration of isolated Mn atoms) should induce intense scattering of carriers, i.e. low mobilities. The latter, according to estimates, should not exceed 10 cm$^2/$(V$\cdot$s), as it is in GaMnAs~\cite{haya1997} and previously studied GaMnSb with highest $T_c$~\cite{rylkov2005}. This contradicts the suggestion that MnSb nanoinclusions are distributed over the whole volume of the film. Thus, a detailed knowledge of samples structure is required to resolve this problem properly. To get this information we performed additional measurements using S/TEM, AFM and MFM methods.

In the present TEM/EDX microanalysis study of the GaSb-MnSb/$\alpha$-Al$_2$O$_3$ system the actual investigation was performed on the cross-sectional piece of sample GM3 with lateral sizes of about 1 $\mu$m. Results are presented in Fig. 4a and 4b.  The film thickness is about 150 nm and the surface roughness does not exceed 6 nm. This is in a good agreement with the data presented in Table 1, which were estimated from the duration of deposition process. The image analysis confirms homogeneous composition throughout the film thickness without any substantial contrast variations. Contrast changes in the lateral direction are due to diffraction contrast arising from the columnar film microstructure which was distinctly observed on bright-field TEM (Fig. 4a) and high-resolution bright-field TEM (HRTEM) images (Fig. 4b), and even effect the HAADF TEM image (not presented here).

\begin{figure}
\centering
\includegraphics[width=6cm]{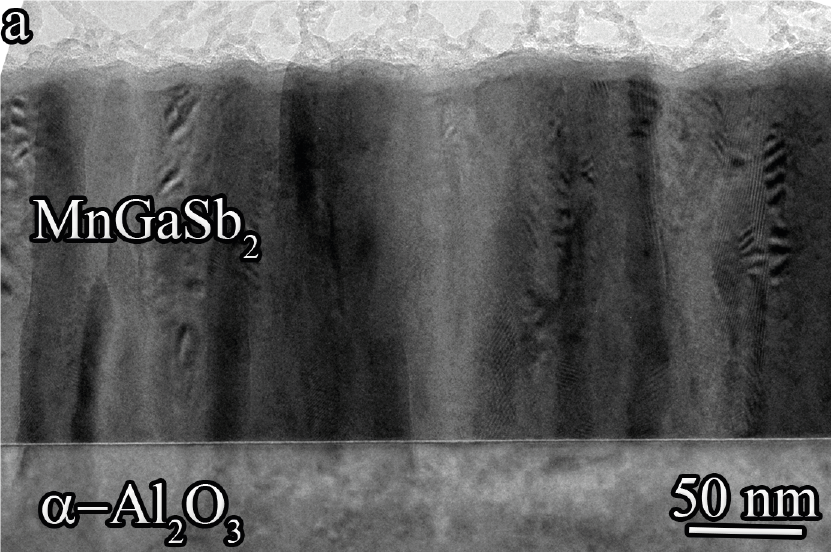}
\\
\includegraphics[width=8.5cm]{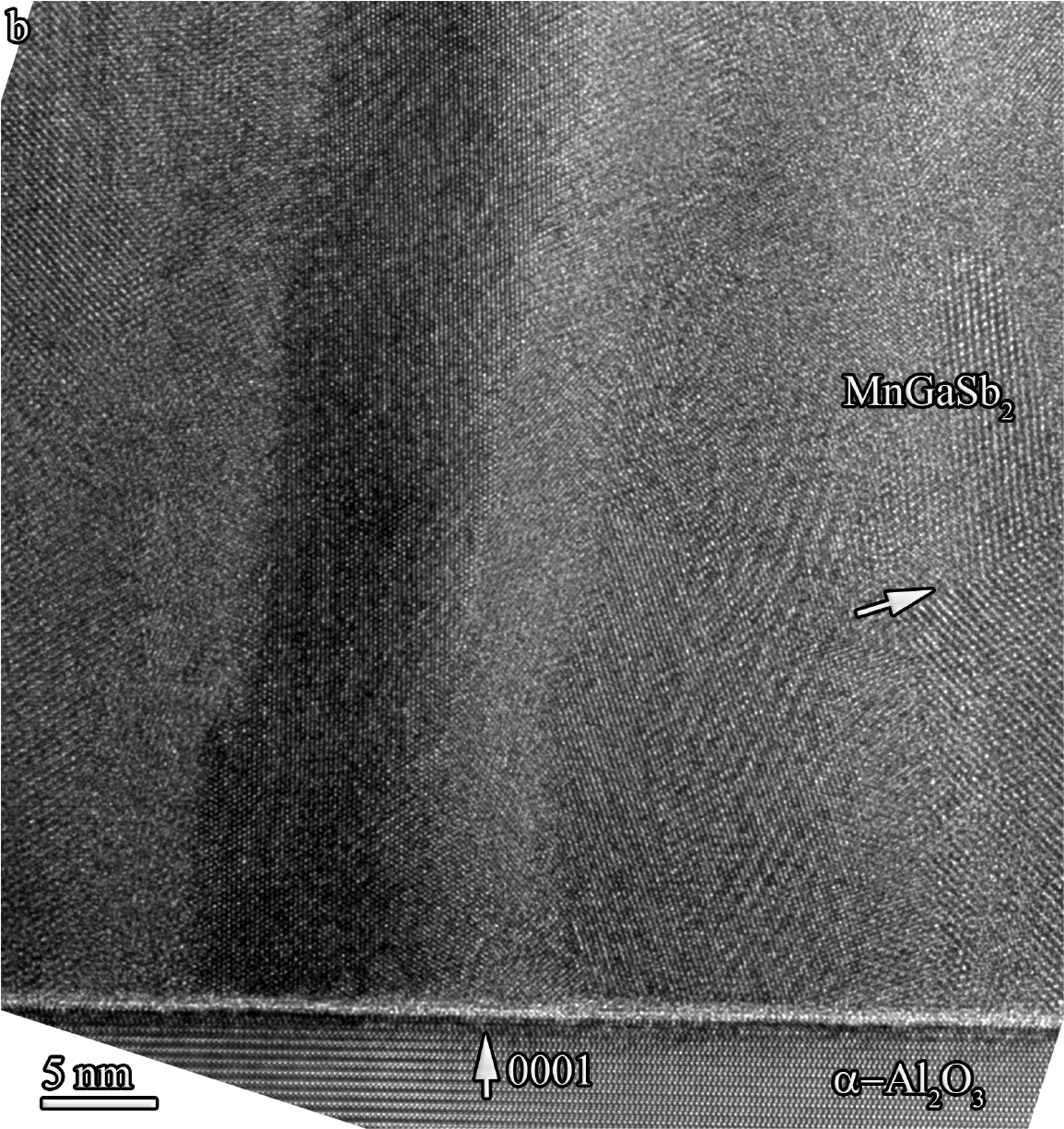}
\caption{TEM images of the film cross-section after annealing (sample GM3): (a) Bright field image. (b) HRTEM image.}
\label{fig5}
\end{figure}

\begin{figure*}[pt]
\centering
\includegraphics[width=4.5cm]{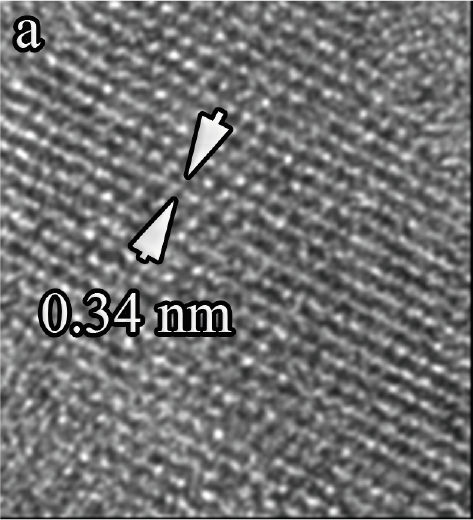}
\includegraphics[width=4.5cm]{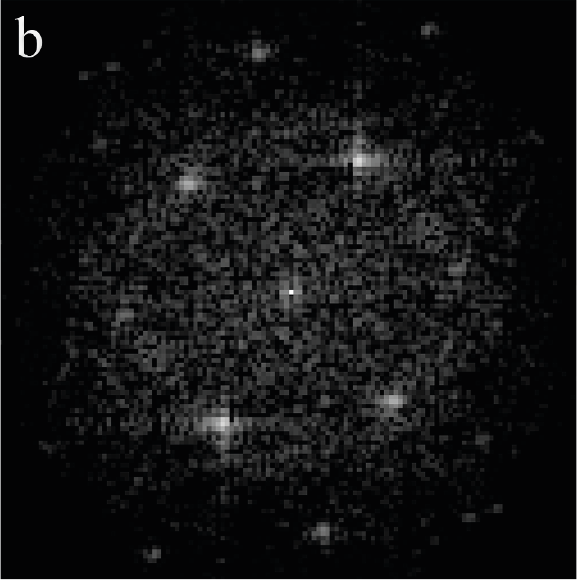}
\includegraphics[width=4.5cm]{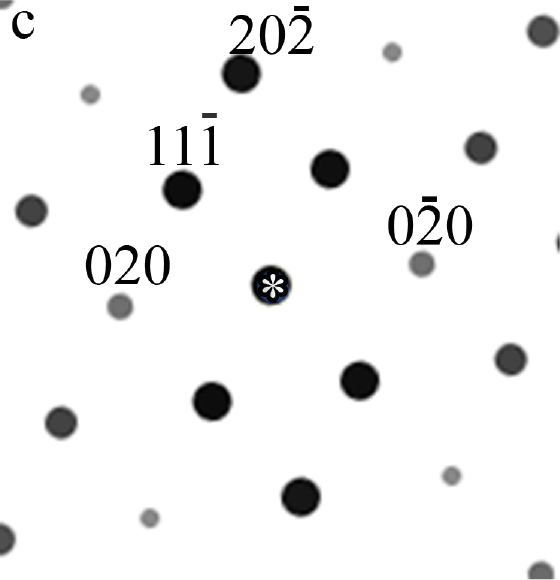}
\includegraphics[width=4.5cm]{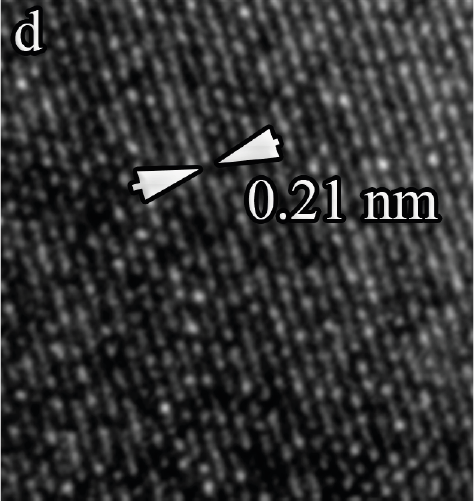}
\includegraphics[width=4.5cm]{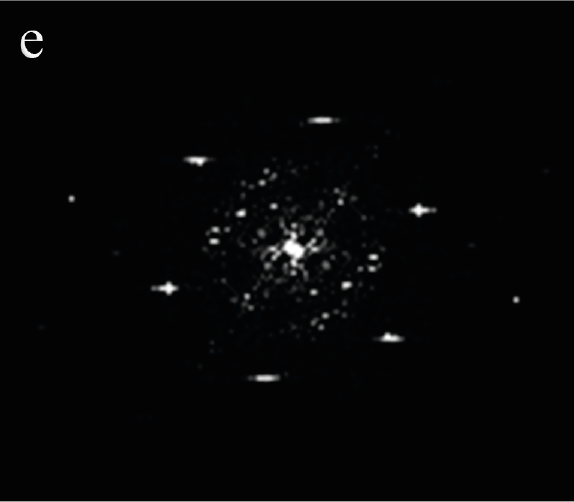}
\includegraphics[width=4.5cm]{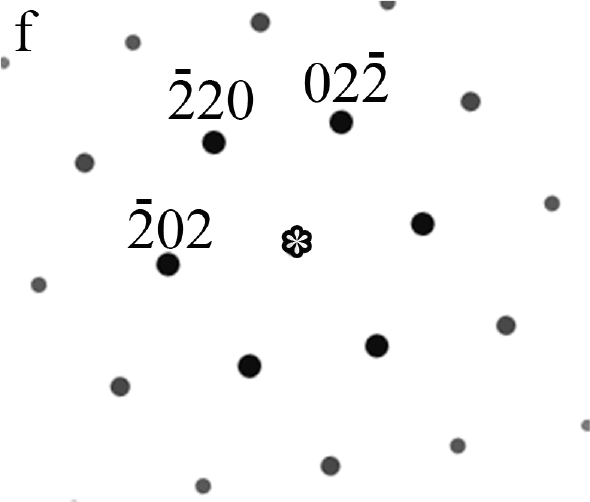}
\caption{(a,d) HRTEM images of sample areas. (b,e) Corresponding two-dimensional Fourier spectra. (c,f) Calculated electronograms of cubic MnGaSb$_2$ compound in [101] (c) and [111] (f) projections.}
\label{fig6}
\end{figure*}

Energy-dispersive X-ray microanalysis (EDX) study of the film composition near the interface edge and at a distance from it yielded the ratio Mn:Ga:Sb=30:30:40 with 2$\%$ accuracy. HRTEM image of studied film is presented in Fig.4b. Fast Fourier Transform (FFT) analysis of high-resolution image areas in two directions (Fig. 5b and 5e) and direct analysis of crystal cell image (Fig. 5c and 5f) were used to analyze the crystalline structure of the sample. As an example, the results for two of ten studied image areas are presented in Fig. 5.

The analysis of all studied areas unambiguously indicate that the structure of the film is cubic and corresponds to the structural type of the GaSb crystal lattice (\emph{space group} $F\overline{4}3m$). Moreover, the images clearly show twins and stacking faults, which are typcal for the cubic GaSb crystals. At the same time, the electron diffraction data obtained for pristine films correspond to diffraction patterns for the hexagonal compound with the \emph{space group} $P6^3/mmc$. Also, the image analysis showed that the morphology of the film and the lateral dimensions of the film columns remained the same after annealing. It should be noted, that energy-dispersive X-ray microanalysis showed a slight (2-3$\%$) decrease of Mn content in the film volume after annealing. This can be related to the details of sample preparation (e.g. the presence of a thicker damaged layer on the surface) or to the diffusion of Mn atoms. Thus, we can conclude that film annealing causes a phase transition of the hexagonal GaSb matrix to cubic one.
\begin{figure}
\centering
\includegraphics[width=4.1cm]{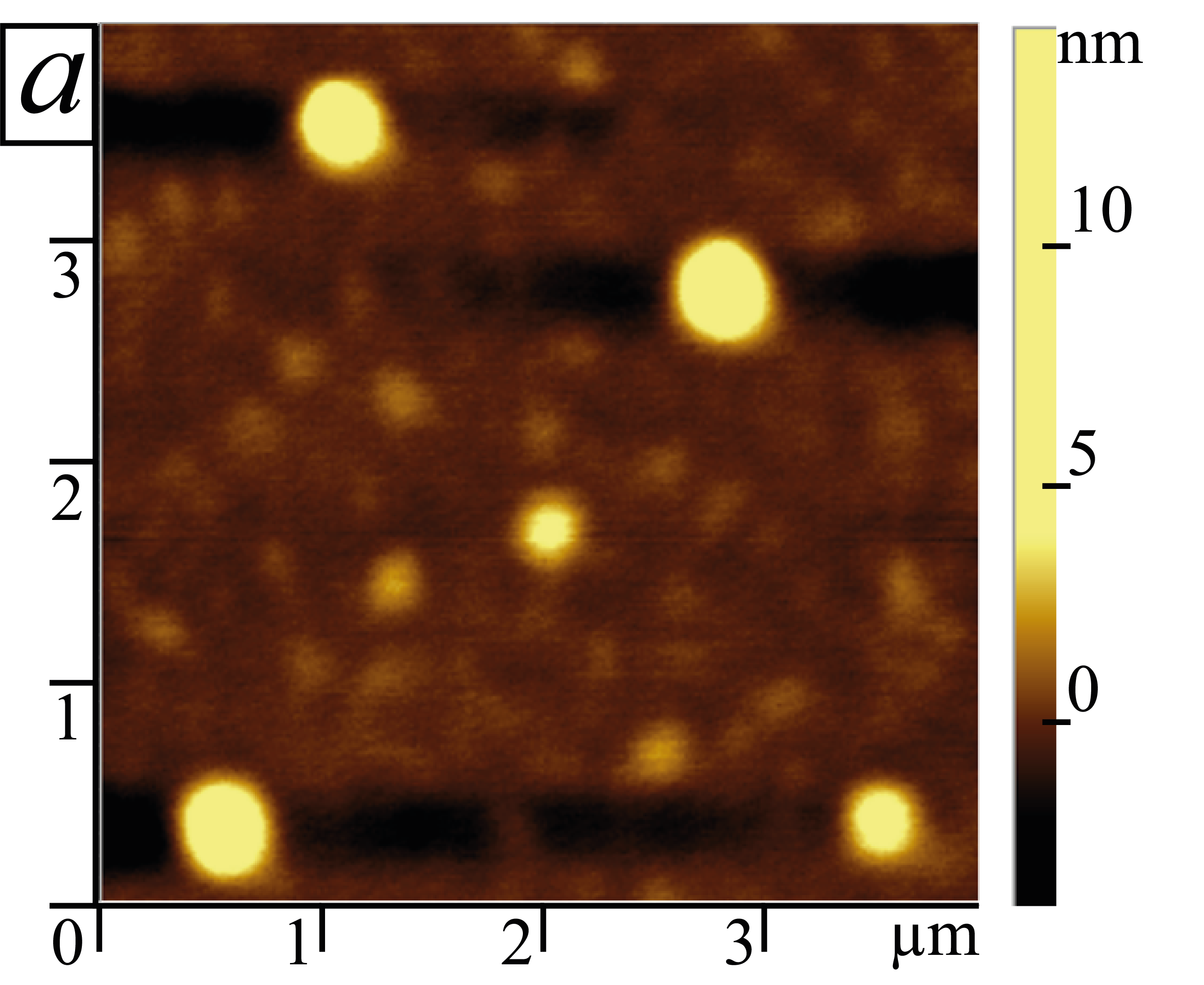}
\includegraphics[width=4.1cm]{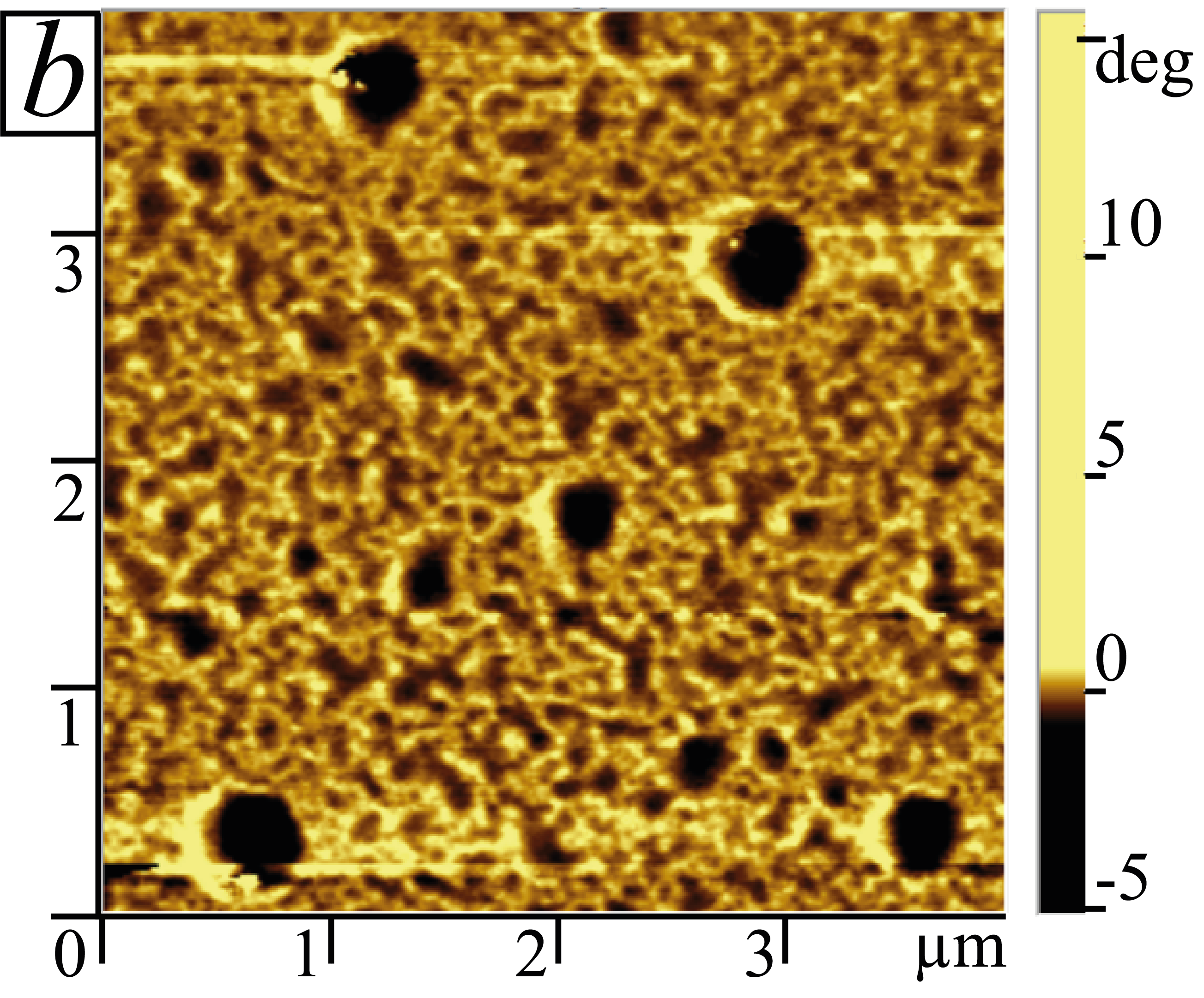}
\includegraphics[width=4.1cm]{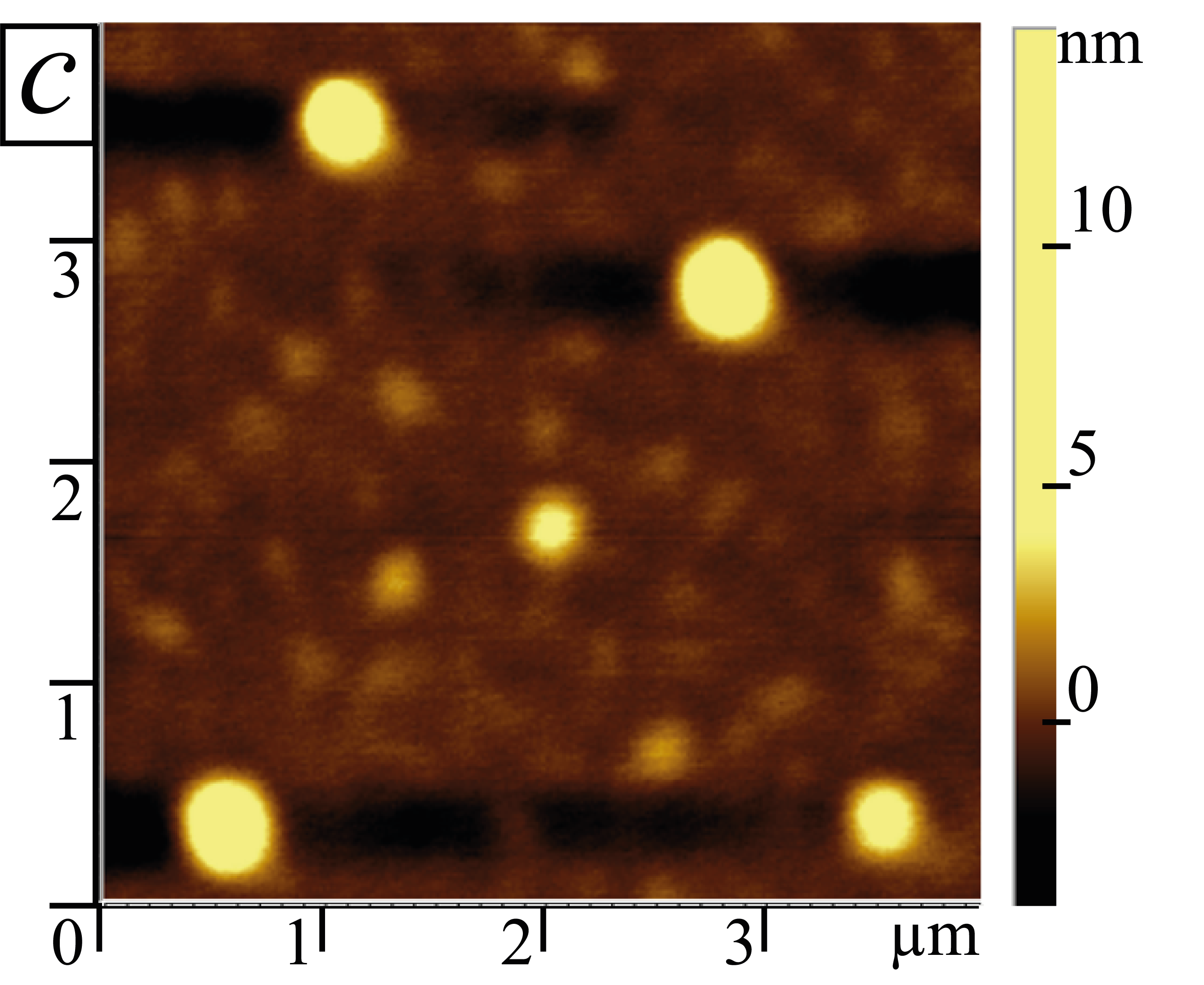}
\includegraphics[width=4.1cm]{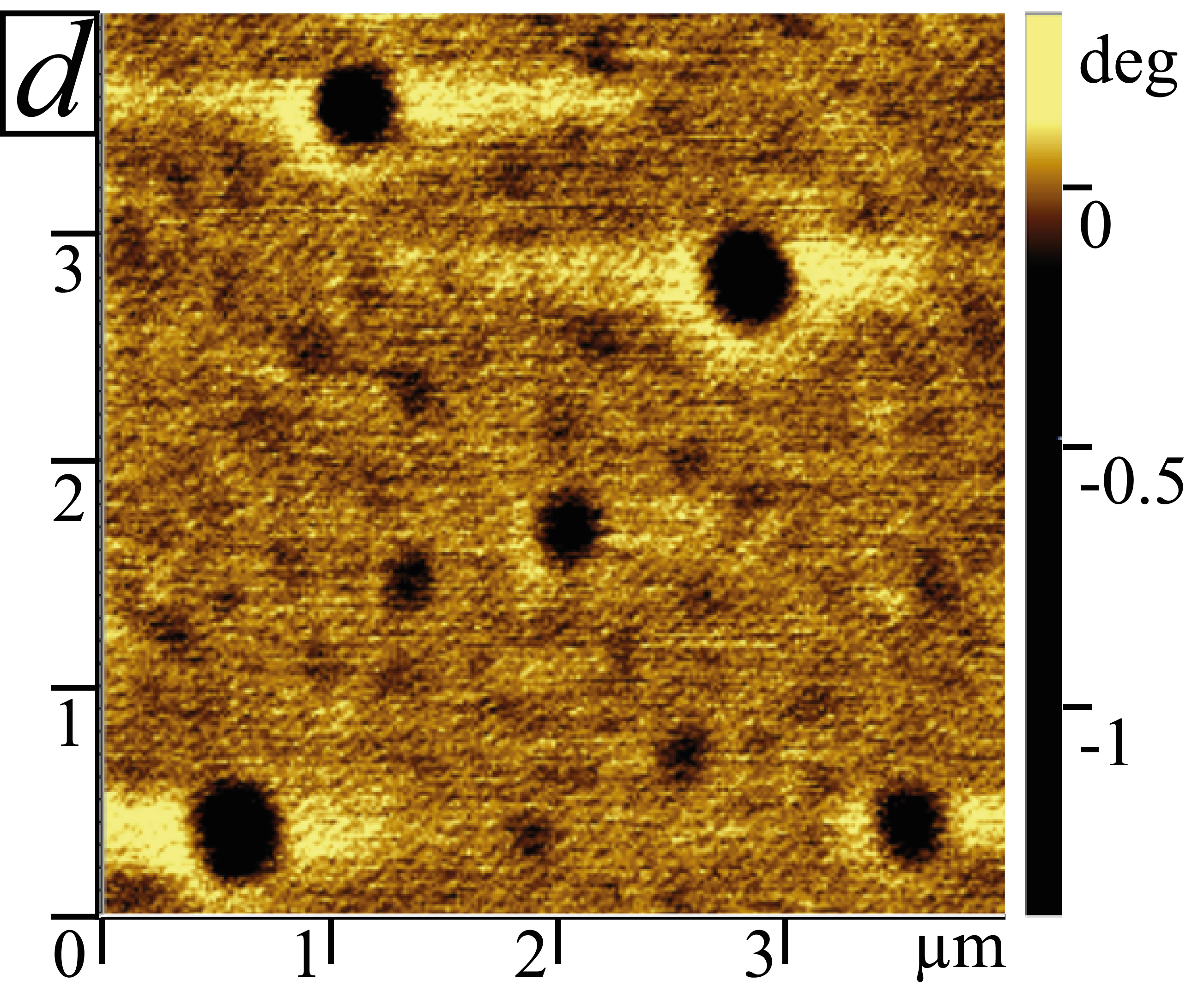}
\caption{(colour online) AFM (a,c) and MFM (b,d) images of the same surface at 303 K (a,b) and 413 K (c,d).}
\label{fig7}
\end{figure}
As a result, the electron microscopy data unambiguously shows that the columnar microstructure of the film persists up to almost surface layers. It should be noted, that no signs of second phase precipitates in the film volume was observed. In particular, using a GaSb-MnSb eutectic composition as a target for sputtering allowed us to avoid MnGa inclusion formation, which were observed earlier in samples obtained by laser deposition from Mn and GaSb targets, without taking the stoichiometry into account~\cite{bobr2013}.

However, due to unevenness of the film and oxidation, the top layer of the film could not be studied carefully and we cannot determine the exact morphology and composition of the surface. So the AFM and MFM measurements were performed.

AFM and MFM images of one sample obtained at 303 and 413 K are presented in Fig. 6. The MFM images clearly show that magnetic properties of the system remain stable at these temperatures, i.e. the sample exhibits ferromagnetism with $T_c$ much higher than room temperature. A comparison of the AFM and MFM data shows that magnetic moments are distributed mostly within areas of larger height (MnSb inclusions) detected by AFM. Combining these results with SEM and TEM data suggests that MnSb inclusions are ferromagnetic and located mostly close to the film surface, rather than evenly distributed within it's volume. This assumption also provides a good explanation of high mobility values of conducting holes, since magnetic inclusions localized on the surface have a substantially weaker effect on scattering in studied films.

\section{Conclusions}
We have studied structural, magnetic and magnetotransport properties of annealed (GaSb)$_{1-x}$(MnSb)$_x$ films with $x=0.41$ and thickness $d = 120-135$ nm. Electron microscopy data suggests that studied films have single-phase columnar structure in it's volume. Moreover, it was found that annealing process induce phase transition of hexagonal GaSb matrix into cubic one. On the other hand, AFM and MFM studies revealed the presence of ferromagnetic MnSb inclusions near surface of the films. This perfectly explains high mobility values of charge carriers in these systems. Experimental observation of these inclusions is crucial for the explanation of high temperature ferromagnetism, since magnetization hysteresis at room temperature perfectly fits the idea of dominant contribution of MnSb inclusions with $T_c > 400$ K interacting via holes. It is confirmed by the appearance of substantial anomalous Hall effect even at room temperature. The interaction of charge carriers with inclusions is greatly affected by the Shottky barriers on GaSb/MnSb boundaries. Due to high carrier concentrations in studied films, the transparency of these barriers is rather high which explains the saturation of magnetization at $p=10^{20}$ cm$^{-3}$.

\section{Acknowledgements}
This work was partially supported by Russian Foundation for Basic Research (grants \#17-02-00262 and \#16-03-00150) and by Ministry of Education and Science of Russian Federation (grant \#16.2814.2017/PCh).


\begin{thebibliography}{}
\bibitem{jung2006}
Jungwirth T, Sinova J, Masek J, Kucera J, MacDonald A H 2006 \emph{Rev. Mod. Phys.} \textbf{78} 809-864

\bibitem{dietl2007}
Dietl T 2007 \emph{Lecture Notes on Semiconductor Spintronics} vol. 712 (Berlin: Springer) pp 1-46

\bibitem{dietl2014}
Dietl T, Ohno H 2014 \emph{Rev. Mod. Phys.} \textbf{86} 187-251

\bibitem{chen2009}
Chen L, Yan S, Xu P F, Lu J, Wang W Z, Deng J J, Qian X,Ji Y, Zhao J H 2009 \emph{Appl. Phys. Lett.} \textbf{95(18)} 182505

\bibitem{wang2014}
Wang M, Marshall R A, Edmonds K W, Rushforth A W, Campion R P, Gallagher B L 2014 \emph{Appl. Phys. Lett.} \textbf{104} 132406

\bibitem{chen2011}
Chen L, Yang X, Yang F, Zhao J, Misuraca J, Xiong P, von Molnar S 2011 \emph{Nano Lett.} \textbf{11(7)} 2584-2589

\bibitem{lawnPSS2011}
Lawniczak-Jablonska K, Libera J, Wolska A, Klepka M T, Dluzewski P, Sadowski J, Wasik D, Twardowski A, Kwiatkowski A, Sato K 2011 \emph{Phys. Status Solidi RRL} \textbf{5} 62-64

\bibitem{lawnJAP2011}
Lawniczak-Jablonska K, Wolska A, Klepka M T, Kret S, Gosk J, Twardowski A, Wasik D, Kwiatkowski A, Kurowska B, Kowalski B J, Sadowski J 2011 \emph{J. Appl. Phys.} \textbf{109} 074308

\bibitem{rylkov2005}
Rylkov V V, Aronzon B A, Danilov Yu A, Drozdov Yu N, Lesnikov V P, Maslakov K I, Podol'skii V V 2005 \emph{J. Exp. Theor. Phys.} \textbf{100(4)} 742

\bibitem{braun2007}
Braun W, Trampert A, Kaganer V M, Jenichen B, Satapathy D K, Ploog K H 2007 \emph{J. of Cryst. Growth} \textbf{301} 50-53

\bibitem{maren2014}
Marenkin S F, Novodvorsky O A, Shorokhova A V, Davydov A B, Aronzon B A, Kochura A V, Fedorchenko I V, Khramova O D, Timofeev A V 2014 \emph{Inorg. Mater.} \textbf{50(9)} 897

\bibitem{kopl2015}
Koplak O V, Polyakov A A, Davydov A B, Morgunov R B, Talantsev A D, Kochura A V, Fedorchenko I V, Novodvorskii O A, Shorokhova A V, Aronzon B A 2015 \emph{J. Exp. Theor. Phys.} \textbf{120(6)} 1012-1018

\bibitem{talan2016}
Talantsev A, Koplak O, Morgunov R 2016 \emph{Superlattices and Microstructures} \textbf{95} 14-23

\bibitem{mats2000}
Matsukura F, Abe E, Ohno H 2000 \emph{J. Appl. Phys.} \textbf{87(9)} 6442-6444

\bibitem{lotin2010}
Lotin A A, Novodvorsky O A, Khaydukov E V, Rocheva V V, Khramova O D, Panchenko V Ya, Wenzel C, Trumpaicka N, Shcherbachev K D 2010 \emph{Semiconductors} \textbf{44} 246

\bibitem{stad1987}
Stadelmann P A 1987 \emph{Ultramicrocopy} \textbf{21} 131

\bibitem{novo2011}
Novotortsev V M, Kochura A V, Marenkin S F, Fedorchenko I V, Drogunov S V, Lashkul A, Lahderanta E 2011 \emph{Russian Journal of Inorganic Chemistry} \textbf{56(12)} 1951-1956

\bibitem{tera1968}
Teramoto I, Van Run A M J G 1968 \emph{J. Phys. Chem. Solids.} \textbf{29} 347-352

\bibitem{M6}
Yakovleva E I, Oveshnikov L N, Kochura A V, Lisunov K G, Lahderanta E, Aronzon B A 2016 \emph{J. Exp. Theor. Phys. Lett.} \textbf{101(2)} 130-135

\bibitem{sze1981}
 Sze S 1981 \emph{Physics of Semiconductor Devices} 2nd ed. (New York: Wiley) p 880

\bibitem{coeh1985}
Coehoorn R, Haas C, Groot R A 1985 \emph{Phys. Rev. B} \textbf{31} 1980

\bibitem{nagaosa}
Nagaosa N, Sinova J, Onoda S, MacDonald A H, Ong N P 2010 \emph{Rev. Mod. Phys.} {\bf 82} 1539

\bibitem{xiao1992}
Xiao  J Q, Jiang J S, Chien C L 1992 \emph{Phys. Rev. Lett.} {\bf 68(25)} 3749-3752

\bibitem{M12}
Oveshnikov L N, Kulbachinskii V A, Davydov A B, Aronzon B A, Rozhansky I V, Averkiev N S, Kugel K I, Tripathi V 2015 \emph{Scientific Reports} \textbf{5} 17158.

\bibitem{M16}
Oveshnikov L N, Nekhaeva E I 2017 \emph{Semiconductors} \textbf{51(10)} 1313-1320

\bibitem{shen2008}
Shen S, Liu X, Ge Z, Furdyna J K, Dobrowolska M, Jaroszynski J 2008 \emph{J. Appl. Phys.} \textbf{103(7)} 07D134

\bibitem{haya1997}
Hayashi T, Tanaka M, Nishinaga T, Shimada H 1997 \emph{J. Appl. Phys.} \textbf{81(8)} 4865-4867

\bibitem{bobr2013}
Bobrov A I, Pavlova E D, Kudrin A V, Malekhonova N V 2013 \emph{Semiconductors} \textbf{47(12)} 1587-1590


% etc
\end{thebibliography}
\end{document}